\documentclass[12pt]{article}
\usepackage{amssymb,epsf}
\def\lsim{\mathrel{\rlap {\raise.5ex\hbox{$ < $}}
{\lower.5ex\hbox{$\sim$}}}}
\def\gsim{\mathrel{\rlap {\raise.5ex\hbox{$ > $}}
{\lower.5ex\hbox{$\sim$}}}}
\def\sqr#1#2{{\vcenter{\vbox{\hrule height.#2pt
        \hbox{\vrule width.#2pt height#1pt \kern#1pt
           \vrule width.#2pt}
        \hrule height.#2pt}}}}

\def\lsim{{\displaystyle
{{\raise-8pt\hbox{$ <$}}
\atop{\raise5pt\hbox{$\sim$}}}}}
\def\gsim{{\displaystyle
{{\raise-8pt\hbox{$ >$}}
\atop{\raise5pt\hbox{$\sim$}}}}}
\def\slsim{{\displaystyle
{{\raise-8pt\hbox{$\scriptstyle <$}}
\atop{\raise5pt\hbox{$\scriptstyle \sim$}}}}}
\def\sgsim{{\displaystyle
{{\raise-8pt\hbox{$\scriptstyle  >$}}
\atop{\raise5pt\hbox{$\scriptstyle \sim$}}}}}
\newskip\humongous \humongous=0pt plus 1000pt minus 1000pt

\newcommand{\sumpf}[0]{\sum_{(H^{\rm f},G^{\rm f})}\! \! \! \!
{\raise
4pt
\hbox{$'$}}\,}
\newcommand{\sump}[0]{\sum_{(H,G)}\! \! {\raise 4pt \hbox{$'$}}\,}

\def\bs{\begin{subequations}}
\def\es{\end{subequations}}
\catcode`\@=11
\newcount\hour
\newcount\minute
\newtoks\amorpm
\hour=\time\divide\hour by60
\minute=\time{\multiply\hour by60 \global\advance\minute by-\hour}
\edef\standardtime{{\ifnum\hour<12 \global\amorpm={am}%
        \else\global\amorpm={pm}\advance\hour by-12 \fi
        \ifnum\hour=0 \hour=12 \fi
        \number\hour:\ifnum\minute<10 0\fi\number\minute\the\amorpm}}
\edef\militarytime{\number\hour:\ifnum\minute<10 0\fi\number\minute}
\def\draftlabel#1{{\@bsphack\if@filesw {\let\thepage\relax
   \xdef\@gtempa{\write\@auxout{\string
      \newlabel{#1}{{\@currentlabel}{\thepage}}}}}\@gtempa
   \if@nobreak \ifvmode\nobreak\fi\fi\fi\@esphack}
        \gdef\@eqnlabel{#1}}
\def\@eqnlabel{}
\def\@vacuum{}
\def\draftmarginnote#1{\marginpar{\raggedright\scriptsize\tt#1}}
\def\draft{\oddsidemargin -.2truein
        \def\@oddfoot{\sl preliminary draft \hfil
        \rm\thepage\hfil\sl\today\quad\militarytime}
        \let\@evenfoot\@oddfoot \overfullrule 3pt
        \let\label=\draftlabel
        \let\marginnote=\draftmarginnote
   \def\@eqnnum{(\theequation)\rlap{\kern\marginparsep\tt\@eqnlabel}%
\global\let\@eqnlabel\@vacuum}  }
\def\subequations{\refstepcounter{equation}%
  \edef\@savedequation{\the\c@equation}%
  \@stequation=\expandafter{\theequation}
  \edef\@savedtheequation{\the\@stequation}
  \edef\oldtheequation{\theequation}%
  \setcounter{equation}{0}%
  \def\theequation{\oldtheequation\alph{equation}}}

\def\endsubequations{\setcounter{equation}{\@savedequation}%
  \@stequation=\expandafter{\@savedtheequation}%
  \edef\theequation{\the\@stequation}\global\@ignoretrue
  \vspace*{-12pt} \\}
\def\bs{\begin{subequations}}
\def\es{\end{subequations}}
\relax
\def\Im{\,{\rm Im}\, }

\def\thefootnote{\fnsymbol{footnote}}
\def\be{\begin{equation}}
\def\ee{\end{equation}}
\def\ba{\begin{eqnarray}}
\def\ea{\end{eqnarray}}

\def\th{\vartheta}

\newcommand{\ar}[2]{{#1\atopwithdelims[]#2}}

\def\ee{\end{equation}}
\def\bea{\begin{eqnarray}}
\def\eea{\end{eqnarray}}
\def\nn{\nonumber}

\def\np#1#2#3{Nucl. Phys. {\bf{B#1}} (#2) #3}
\def\pl#1#2#3{Phys. Lett. {\bf{B#1}} (#2) #3}

\def\prl#1#2#3{Phys. Rev. Lett. {\bf{#1}} (#2) #3}
\def\pr#1#2#3{Phys. Rev. {\bf{D#1}} (#2) #3}

\newcommand{\uarrw}[0]{\mathrel{
{\raise.5ex\vbox{\hrule width 1cm}\hskip-6pt\rightarrow}}}
\def\thebibliography#1{%
\vskip 0.5cm \centerline{\bf References}
\list{%
[\arabic{enumi}]}{\settowidth\labelwidth{[#1]}
\leftmargin\labelwidth
\advance\leftmargin\labelsep
\usecounter{enumi}}
\def\newblock{\hskip .11em plus .33em minus .07em}
\sloppy\clubpenalty4000\widowpenalty4000
\sfcode`\.=1000\relax}

\renewcommand{\theequation}{\arabic{section}.\arabic{equation}}

\renewcommand{\section}{\setcounter{equation}{0}\@startsection%
{section}{1}{0mm}{-\baselineskip}{0.5\baselineskip}%
{\normalfont\normalsize\bfseries}}

\renewcommand{\subsection}{\@startsection%
{subsection}{2}{0mm}{-\baselineskip}{0.5\baselineskip}%
{\normalfont\normalsize\slshape}}
\begin{document}
\renewcommand{\theequation}{\arabic{section}.\arabic{equation}}
\begin{titlepage}
\begin{flushright}
Bicocca-FT/00/05,\\
hep-th/0004070 
\end{flushright}
\begin{centering}
\vspace{1.0in}

{\bf \large type IIB orbifolds with D5-branes
and their string duals}\ $^\dag$
\\
\vspace{1.7 cm}
{\bf { Andrea Gregori$^1$}} \\
\medskip
\vspace{.4in}
{\it  Dipartimento di Fisica, Universit{\`a} di Milano--Bicocca}\\
{\it and}\\
{\it INFN, Sezione di Milano, Italy}\\

\vspace{2.5cm}
{\bf Abstract}\\
\vspace{.1in}
We consider $Z_2$, freely acting orbifolds of the type IIB string
with 16 parallel D5-branes.
When the string is compactified on $T^2 \times T^4$ and 
the D5-branes are wrapped on $T^2$, these systems possess
${\cal N}_4=2$ supersymmetry, originating from the spontaneous, partial
breaking of ${\cal N}_4=4$. 
Extended supersymmetry allows us to investigate the duality
with certain heterotic and type I constructions, and to obtain
informations about their non-perturbative regime. 
\end{centering}
\vspace{4cm}

\hrule width 6.7cm
$^\dag$\  Research partially supported by the EEC under the contract\\
TMR-ERBFMRX-CT96-0045.\\
\\
$^1$e-mail: agregori@pcteor.mi.infn.it

\end{titlepage}
\newpage
\setcounter{footnote}{0}
\renewcommand{\thefootnote}{\arabic{footnote}}

\setcounter{section}{1}
\section*{\normalsize{\bf 1. Introduction}}

Systems of type IIB strings in the presence of $N$ parallel Dp-branes
are of much interest for the investigation of the non-perturbative 
aspects of Yang-Mills theories. A well studied situation is 
when one takes the limit of large $N$, small $g_s$.
In this limit, gravity is effectively decoupled,
and one reduces to study a pure gauge theory.
Recently, situations in which the string coordinates transverse to the 
D-branes enter in the game have also been considered \cite{bkl}.
In this work, we will only consider BPS D-branes;
the presence of such D-branes leads to the
breaking of half of the supersymmetries of the type IIB string.
When the four coordinates transverse to the branes are toroidally compactified,
cancellation of the Ramond-Ramond charge requires the introduction
of orientifold O5 planes and fixes the number of 
D5-branes to be sixteen, toghether with their mirror pairs.
This construction can be seen as an orientifold of the type IIB string,
projected by $\tilde{\Omega} \equiv \Omega \times I_{(7,8,9,10)}$,
the product of the world-sheet parity times the target-space reflection
acting on the transverse coordinates.
This orbifold is therefore related to the ordinary orientifolds
by T-duality along $x_7$, $x_8$, $x_9$, $x_{10}$, and 
it is reasonable to ask wether this theory is dual to the 
toroidally compactified heterotic string.
The point is non trivial, because, on the type IIB string, supersymmetry
appears as ``spontaneously'' broken, in the sense that
the theory possesses an approximate restoration of the initial 32
supersymmetry charges when there is a local cancellation of the
Ramond-Ramond charge and 
the space transverse to the branes is decompactified. 
On the other hand, it is known that on the heterotic side,
even at the strong coupling there are no more than 16 supercharges: 
the strong coupling limit is in fact the type IIA string compactified on K3.
In the case of heterotic compactifications with maximal supersymmetry,
such a duality can therefore exist only in a particular limit
in the moduli space of the two theories. It is however possible that
it exists for theories with a  lower number of 
supersymmetries, in which the projection that breaks supersymmetry
has a non-trivial action on the dilaton field. 
In order to understand the situation,
we compactify two further coordinates on $T^2$ and 
consider the theory obtained by adding a freely acting,
$Z_2$ projection that acts as a twist on the transverse coordinates
and as a translation along a circle of the two-torus parallel to the branes. 
This projection breaks spontaneously the supersymmetry to ${\cal N}_4=2$:
the ${\cal N}_4=2$ and ${\cal N}_4=4$ phases are then continuosly related.
In  the ${\cal N}_4=2$ phase of the theory,
the moduli dependence of certain amplitudes is reacher than in the 
${\cal N}_4=4$ phase. 
Supersymmetry is nevertheless still sufficiently extended to allow
the comparison of amplitudes that, like the $R^2$ term we will consider,
receive a contribution only from non-perturbatively stable, BPS states.

When $V_{(4)}$, the volume of the space transverse to the D5-branes, is small,
the ordinary type I orientifold description is indeed the more appropriate.
The description in terms of D5-branes turns out to be useful in order
to investigate the limit $V_{(4)} \to \infty$,
non-perturbative from the type I point of view.
As we will discuss, in this limit there is an approximate
restoration of a larger amount of supersymmetry. Due to the free action
of the orbifold projection, in this model indeed act both a perturbative
and a non-perturbative super-Higgs mechanism, responsible
for the spontaneous breaking of the ${\cal N}_4=8$ supersymmetry.
We argue that this model, T-dual to a type I freely acting orbifold 
with sixteen D9-branes, presented in Ref. \cite{adds},
is also dual to a freely acting
orbifold of the heterotic string, with gauge group of maximal rank.
In order to substantiate our arguments, we consider the renormalization of the
effective coupling of the $R^2$ term.
We compare this quantity, as computed on the heterotic side, 
on the type I and in the ``type IIB'', $\tilde{\Omega}$ 
dual orientifold, both for $V_{(4)} \to 0$ and $V_{(4)} \to \infty$.
Since the effective coupling of this term depends on the moduli
of the vector multiplets, this comparison 
provides us with informations about the non-perturbative behavior
of the theory.

The paper is organized as follows:

In Section 2 we discuss the type IIB orientifold with D5-branes, 
and the further breaking of supersymmetry
produced by the orbifold projection. The main part of the section is devoted
to the discussion of the gravitational corrections. 

In Section 3 we present the heterotic and type I duals. 
We compare then the gravitational corrections as computed in all
the constructions and discuss the map relating the moduli in the vector 
manifolds of the effective theories.

In Section 4 we review the procedure of 
rank reduction in type I strings. This was first discussed
in \cite{bps}--\cite{a}, and we apply it to the case of type I models
with spontaneous breaking of supersymmetry.
We use then the results of the previous sections in order to
infer the non-perturbative behavior of these constructions.

Further comments and conclusions are in Section 5.

\noindent

\vskip 0.3cm
\setcounter{section}{2}
\setcounter{equation}{0}
\section*{\normalsize{\bf 2. The type IIB string with D5-branes }}

We start by considering the addition of parallel D5-branes to the
type IIB superstring. When the space transverse to the D5-branes
is compactified on $T^4$, the Ramond-Ramond charge of the branes
must be cancelled. The cancellation is obtained by introducing orientifold
O5 planes. By a chain of T-dualities applied to the O9 case, 
one can see that their number is fixed to sixteen: 
the allowed number of D5-branes
is therefore also sixteen (toghether with their sixteen mirrors).
If each mirror pair sits on one orientifold plane, the cancellation
of the RR charge is local.
This construction can be seen as an orientifold obtained by
projecting the type IIB string with $\tilde{\Omega}$,
defined as the product $\Omega \times I_{(7,8,9,10)}$,
of the world-sheet parity, $\Omega$, and
the target-space reflection $I_x : \; x \to - x$, along the four
transverse coordinates. 
T-duality on the type IIB string, along these coordinates,
relates therefore this theory to the ordinary type I, orientifold
construction, in which the D9-branes and one O9 plane replace
the D5-branes and the O5-planes. When the volume, $V_{(4)}$, of the space
transverse to the D5-branes, is small, the type I orientifold
indeed gives the more appropriate description of the theory,
that corresponds to the phase in which the D-branes gauge group is broken
to $U(1)^{16}$.
The opposite limit, $V_{(4)} \to \infty$, is non-perturbative from the point
of view of the effective theory.
On the other hand, a simple look at the orientifold construction
indicates that, since the $\tilde{\Omega}$ projection breaks
T-duality along $T^4$, this limit is indeed very different from
the other one: 
in this limit the bulk theory feels the D5-branes very weakly, and we
expect that, being the  supersymmetry breaking projection
associated to the breaking of T-duality, to recover an
approximate restoration of the original amount of supersymmetry, that appears
therefore spontaneously broken for any finite value of $V_{(4)}$. 
It is not difficult to write an explicit expression for the partition 
function of this kind of orientifold: it can be obtained from the
well known orientifold partition function with D5-branes, by
performing a translation along the four circles transverse to the branes,
up to the center of the segments connecting the two fixed points
of each circle. In the closed string sector, the Torus, being
invariant under translations, is not affected.
The Klein bottle and the open string diagrams are instead affected;
in these sectors the translation amounts to a half-integer shift
of the winding numbers corresponding to the transverse coordinates.
When the radii are small ($V_{(4)} \to 0$), the theory recovers
the ordinary, ``unshifted'' orientifold. In the large radii limit
instead ($V_{(4)} \to \infty$), the diagrams generated by the orientifold
projection vanish, leaving only the type IIB partition function.

We now go to four dimensions by compactifying two other 
coordinates on a two-torus, and break the ${\cal N}_4=4$ supersymmetry
with a freely acting, $Z_2$ orbifold projection, that acts
as a twist, $x \to - x$, on $x_7$, $x_8$, $x_9$, $x_{10}$,
and as a half circumference translation in a circle of $T^2$.
This produces a spontaneous breaking of supersymmetry to ${\cal N}_4=2$
\cite{kk}, due to a stringy version of the mechanism of Ref. \cite{scsc}.
This model is related by T-duality along $T^4$ to the
``Scherk--Schwarz breaking'' type I construction presented in 
Ref.~\cite{adds}. The massless spectrum is therefore the same.
Here however, having a local cancellation of the Ramond-Ramond 
charge, necessary for the existence of a regular decompactification
limit, requires to put each D5-branes pair on its corresponding orientifold
plane. This leads to the breaking of the gauge group to $U(1)^{16}$.
This configuration corresponds to introducing appropriate
Wilson lines on the dual construction with D9-branes.
Under T-duality, Wilson lines are in fact mapped into
the positions of the D5-branes.  

\subsection*{The $R^2$ corrections}
 
We come to the computation of the corrections to the effective
coupling of the $R^2$ term\footnote{For a precise definition and a discussion
of this amplitude in our notation, we refer the reader to \cite{6auth}.}.
We are interested in the expression of the effective coupling
in the two limits of large and small $V_{(4)}$. As we observed,
for small $V_{(4)}$, the theory is perturbatively equivalent to
a type I orientifold, with D9-branes instead of D5-branes.
The computation of the gravitational corrections is therefore
absolutely equivalent, modulo a redefinition of the fields in the
vector manifold, due to the D5-/ D9-branes exchange, to the one
already performed in that framework. 
Taking this into account, we can immediately
write the correction to the effective coupling by appropriately
adapting the result obtained for that model in Ref. \cite{gk}
(see also Section 3):
\be
{16 \, \pi^2 \over g^2_{\rm   grav}(\mu )}
\, =  \, 16 \, \pi^2 \Im S^{\prime}
-{2 \over 3} \log \Im U \left \vert \vartheta_4 \left( U \right) \right\vert^4
- b \log {\mu  \big/ M } \, ,
\label{5branes}
\ee
where $U$ is the modulus associated to the complex structure of the 
unwisted two-torus and
now $\langle \Im S^{\prime} \rangle$ is the tree level coupling
of the D5-branes \cite{green}, 
expressed as the vacuum expectation value  of the field 
$S^{\prime}$, defined as in Ref. \cite{par} (see
also Eq. (\ref{sp}) below).
$2 \big/ 3$ is the value of the beta-function coefficient, 
$b_{\rm grav}={24 -N_V + N_H \over 12}$. 
For the particular parametrization of the untwisted two-torus in which
\be
\Im T = R_5 R_6,~~~~~~~~~~~~\Im U= R_6 / R_5 \, ,
\label{turr}
\ee
the $\vartheta_4 (U)$ appears  when the orbifold projection that twists 
the four coordinates transverse to the branes acts on the two-torus 
as $(-1)^m$, on the momenta on $x_6$.
The third term on the r.h.s. contains
the infrared running, as a function of the infrared cut-off
$\mu$ and the characteristic mass scale $M$.
The ratio of these quantities is invariant under type II/heterotic/type I,
string-string dualities, and can be expressed as the ratio
of a physical cut off $\sqrt{p^2}$ and the Planck mass $M_P$.
The $\beta$-function coefficient $b$, in the cases it can 
be explicitly computed
(see for instance \cite{gkp}), turns out to be
$b={B_4 - B_2 \over 3}$, with $B_2$ and $B_4$ the constant, massless
contributions at a generic point in the moduli space 
(i.e. away from points of enhanced symmetry) to the
second and fourth helicity supertraces (for a discussion of these quantities,
see for instance \cite{kbook}). 

We consider now the limit of (very) large $V_{(4)}$.
As we noticed, in the absence of the orbifold projection
this would be a limit of approximate restoration of the full
supersymmetry. In the present case, this is still true,
provided the circle of the two-torus translated by the orbifold
projection is appropriately decompactified, in order to make vanishing
the orbifold action. In this limit, the correction to the
gravitational coupling is expected to vanish (or, more generally,
to diverge only as a logarithm of the moduli). 
For finite values of the radii
of the two-torus, supersymmetry is instead broken
to ${\cal N}_4=2$. The contribution of the D5-branes is on the
other hand suppressed, and we expect that there is
no tree level contribution to the correction. The full result
comes therefore from one loop (owing to extended supersymmetry,
there are no contributions from higher orders).
The correction to the $R^2$
term is then well approximated by the expression obtained
by inserting the ``helicity operator'' $Q^2 \bar{Q}^2$,
in the type IIB partition function. 
$Q$ and $ \bar{Q}$ are the operators corresponding to the
left and right helicity, and act as derivatives on the
appropriately deformed partition function \footnote{See for instance
Ref. \cite{kbook} for a review.}.
The gravitational amplitude is in fact
proportional to an index (see Refs. \cite{agn,hm}) and, as shown in
Refs. \cite{6auth,infra}, this insertion leads to the 
computation of the appropriate amplitude,   
provided one keeps the
only $U$-dependent part of the result.  
The partition function we use for the  one-loop 
computation is therefore:
\ba
Z_{\rm bulk} & = & { 1 \over \Im \tau \, \vert \eta \vert^{20} } \,
{1 \over 2} \sum_{\alpha,\beta} (-1)^{\alpha + \beta + \alpha \beta} \, 
\theta^2 \ar{\alpha}{\beta} \,
{1 \over 2} \sum_{\bar{\alpha},\bar{\beta}}
(-1)^{ \bar{\alpha} + \bar{\beta} + \bar{\alpha} \bar{\beta} } \, 
\bar{\theta}^2 \ar{\bar{\alpha}}{\bar{\beta}} \nn \\
&& \times ~ {1 \over 2} \sum_{H,G} \, \theta \ar {\alpha+H}{\beta+G}
\theta \ar {\alpha-H}{\beta-G}
\bar{\theta} \ar {\bar{\alpha}+H}{\bar{\beta}+G}
\bar{\theta} \ar {\bar{\alpha}-H}{\bar{\beta}-G} \;
\Gamma_{2,2} \ar{H}{G} \, \tilde{\Gamma}_{4,4} \ar{H}{G} \, .
\label{zbulk}
\ea
We have defined $\tilde{\Gamma}_{4,4} \ar{H}{G}$
as the contribution of the four twisted bosons.
In the ``${\cal N}_4=4$'' sector, namely for $(H,G) \neq (0,0)$,
$\tilde{\Gamma}_{4,4} \ar{H}{G} \equiv {1 \over \eta^4 \bar{\eta}^4}
\Gamma_{4,4} \ar{H}{G}$,
and (\ref{zbulk}) coincides with the partition function of the
${\cal N}_4=4$ freely-acting orbifold of \cite{6auth}. 
In the untwisted, unprojected sector, $(H,G)=(0,0)$, 
it is the usual sum over lattice momenta. 
In the non-compact limit, it is given instead by:
\be
\tilde{\Gamma}_{4,4} \ar{0}{0} \, \equiv \,
 {1 \over \Im \tau^2 \eta^4 \bar{\eta}^4} \, .
\ee
In the lattice contribution of the two-torus, $\Gamma_{2,2}\ar{H}{G}$,
the arguments indicate the shift produced by the translation
associated to the orbifold twist.
The one loop contribution is  
computed  by regularizing the infrared with the 
method of Ref. \cite{infra}.
We obtain:
\be
{ 16 \, \pi^2 \over g^2_{\rm grav}} \approx 
-{2 \over 3} \, \log \Im U \vert \vartheta_4 \left( U \right) \vert^4
- b \log {\mu \big/ M} \, ,
\label{IIbthr}
\ee 
where $\approx$ indicates that in the above expression
we discard terms suppressed at large $V_{(4)}$,
The correction (\ref{IIbthr}) has 
a logarithmic divergence in the limit
$R_6 \to \infty$ ($\Im U \to \infty$), 
where the ${\cal N}_4=8$ supersymmetry is restored. 
This logarithmic dependence can be removed with an appropriate 
choice of the infrared cut-off\footnote{For more details about 
these issues, we refer the reader to 
\cite{6auth,gkp,infra,gkr}--\cite{solving}.}.

\noindent

\vskip 0.3cm
\setcounter{section}{3}
\setcounter{equation}{0}
\section*{\normalsize{\bf 3. The ${\cal N}_4=2$ heterotic and type I duals}}

The heterotic dual is constructed as a $T^2 \times T^4 \big/ Z_2$
orbifold, in which the twist is accompanied by a $Z_2$ translation
on $T^2$, producing a left-right symmetric shift of the momenta;
modular invariance requires then an embedding
of the spin connection into the gauge group.
When the gauge group is broken to $U(1)^{16}$ by Wilson lines, 
the massless spectrum
originating from the currents contains only sixteen vector multiplets
and no hypermultiplets. At this point, it is easy to write the
partition function and to compute the ``regular''
gravitational amplitude, obtained by correcting the ``$\, R^2 \, $'' term
with appropriate gauge amplitude terms.
An example of such a subtraction, valid for orbifolds with
an equal number of vector and hypermultiplets originating from the
$c=16$ currents, $N_V=N_H$, was discussed in Refs. \cite{gk,gkp,gkp2}.

The type I dual was first constructed in Ref. \cite{adds},
as an orientifold of the type IIB string with an ${\cal N}_4=8$ supersymmetry
spontaneously broken to ${\cal N}_4=4$.
The freely acting projection $Z_2^{\rm F}$, that breaks supersymmetry on 
the type IIB ``parent'' string,
acts as a left-right symmetric reflection on four coordinates
of the target space and as a half-circumference translation
in another circle. The translation is effectively introduced
by a projection onto the quantum numbers of the lattice
of momenta associated to that compact coordinate.
Indeed, in Ref. \cite{adds} two such constructions were discussed.
They are distinguished by the action of this projection on this lattice:
either as ${ 1+(-1)^{m} \over 2}$
or as ${ 1+(-1)^{n} \over 2}$. One gets type I orientifolds 
with different properties: in the first case, the ``momentum shift'',
one gets a model with gauge group $SO(16)$, entirely
provided by open string states ending on D9-branes:
no D5-branes are present. This model, called ``Scherk-Schwarz breaking''
in Ref. \cite{adds}, is dual to the heterotic construction. 
In order to obtain hypermultiplets from the
the D-branes massless spectrum, it is necessary to introduce
a Wilson line, coupled to the
supersymmetry breaking projection. In this case, 
the gauge group splits into factors $SO(N_1) \times SO(N_2)$,
with hypermultiplets in the $({\bf N_1}, {\bf N_2})$, bifundamental
representation. By introducing appropriate 
Wilson lines it is possible to break the gauge group to $U(1)^{16}$,
and obtain the dual of the $\tilde{\Omega}$ orientifold
presented in the previous section.

More subtle is the case of the ``winding shift'', otherwise
called M-theory breaking in \cite{adds}.
In this case, supersymmetry appears to be
unbroken  in the open string sector, that provides
the massless states of a ${\cal N}_4=4$ super Yang-Mills theory
with maximal gauge group $SO(16) \times SO(16)$: the first factor
originates from D9-branes states, while the second factor
originates from a D5-branes sector.
It seems therefore that such construction cannot be dual to the other
models we have considered.
The problem is not constituted by the presence of supersymmetry
in a part of the open string sector, because interactions 
with the other sectors communicate the breaking of supersymmetry also to this 
sector. However, even if at the end the massless spectra
at the $U(1)$ most probably coincide, there is a mismatch between
heterotic and type I constructions: in the latter, the
D9- and D5-branes gauge groups have independent gauge couplings,
while on the heterotic side, the full gauge group has the same coupling:
$g^{-2} \approx \langle  \Im S \rangle$, where $S$ is the heterotic 
dilaton--axion field. There is no loop renormalization, and
non-perturbative corrections cannot produce a linear dependence 
such to dramatically correct the conclusions obtained
by looking at the tree level contribution. We nevertheless keep
on considering here also this construction, because
our arguments, slightly modified, will enable us to 
obtain informations about the non-perturbative regime of this
construction too.

\subsection*{The $R^2$ corrections}

Both in the heterotic and in the type I models,
the gravitational amplitude receives a contribution
only at the tree and one loop levels.
On the heterotic side,
the one loop contribution is easily computed by inserting in the
partition function appropriate operators; their action on the left-movers
is that of the helicity operator, $Q^2$ (see Ref. \cite{infra}). 
After saturation of the fermion zero modes,
their action on the right movers is reduced to a covariant differentiation on
certain bloks of the ``${\cal N}_4=2$'' sector of the partition function,
namely on the second helicity supertrace, $B_2$ \footnote{We refer the reader
to Refs. \cite{infra,kkpr} for more details about these 
computations.}.
This quantity possesses universality properties \cite{kkprn},
and is fixed by the choice of the $Z_2$ shift on $T^2$ and
by the value of the difference, $N_V-N_H$, of vector and hypermultiplets
originating from the currents, in the untwisted sector of the orbifold.
In our case $N_V-N_H=16$, and we have:
\be
B_2 = {1 \over \overline{\eta}^{24}} 
\sump  \Gamma_{2,2} \ar{H}{G}
\overline{\Omega} \ar{H}{G} \, ,
\ee
where the prime indicates that the sum is taken only over the values
$(H,G)$$=\{(0,1)$, $(1,0)$, $(1,1) \}$. 
The modular forms $\Omega \ar{H}{G}$ read:
\ba
\Omega \ar{0}{1}&=&{\hphantom{-}}
\frac{1}{2}
\left(\th_3^4+\th_4^4\right)
\th_3^8\, \th_4^8
\nonumber\\
\Omega \ar{1}{0}&=&-
\frac{1}{2}
\left(\th_2^4+\th_3^4\right)
\th_2^8\, \th_3^8
\label{Om88a}\\
\Omega \ar{1}{1}&=&{\hphantom{-}}
\frac{1}{2}
\left(\th_2^4-\th_4^4\right)
\th_2^8\, \th_4^8 \, .\nn
\ea
The amplitude we need to compute is: 
\be
\langle R^2 \rangle_{(\rm reg)} =
\langle R^2 \rangle 
+ {1 \over 12} \langle P^2 \rangle_{(T^2)} 
+ {5  \over 48} \langle F^2  \rangle_{\rm gauge}  \, . 
\label{rff}
\ee
The first term on the r.h.s. is the ordinary $R^2$ amplitude, computed
with the standard methods (see for instance \cite{agn,infra,kkpr}.
The second and third term 
correspond respectively to the amplitude $F^2$ of the $U(1)^2$ of the 
two-torus, computed as in Ref. \cite{gkp},
and to the gauge amplitude $F^2$ of the currents; the latter 
is obtained by inserting in the one-loop vacuum amplitude the operator:
\be
P^2_{(\rm gauge)} = Q^2 \overline{P}^2_{(\rm gauge)} \, ,
\ee 
where $Q^2$ is the left-helicity operator and
$\overline{P}^2_{(\rm gauge)}$ acts as ${ -1 \over 2 \pi {\rm i}}
{{\cal D} \over {\cal D} \bar{\tau}}$ on the weight eight block
corresponding to the contribution of the $c=16$ lattice.
The correction of the gravitational amplitude provided by the
second and third term in the r.h.s. of (\ref{rff}) was shown
in Ref. \cite{gkp} to be necessary in order to subtract singularities that
don't have a counterpart in dual string constructions, and
to provide therefore a quantity suitable for a comparison
of dual theories. 
In the present case, in the $(H,G) \neq (0,0)$ sector, this is given by:
\be
{\overline{\vartheta}^2 \ar{1+H}{1+G} \overline{\Omega} \ar{H}{G}
\over 4 \overline{\eta}^6} \, .
\ee
Acting on $B_2$ with the right-moving part of (\ref{rff}), 
$\overline{R}^2_{(\rm reg)}$, we obtain:
\be
\overline{R}^2_{(\rm reg)} B_2 \, = \,
{2 \over 3} \, \sump \Gamma_{2,2} \, \ar{H}{G} \, .
\label{rb}
\ee
The final result is given by the infrared-regularized integral
of (\ref{rb}) on the fundamental domain.
On the type I side, the  gravitational corrections
are proportional to an index \cite{ser}, and the
analog of (\ref{rff}) simply amounts to a proper choice of
normalization. The analogous of (\ref{5branes}) and (\ref{IIbthr}) 
for the heterotic model reads therefore:
\ba
{16 \, \pi^2 \over g^2_{\rm   grav}(\mu )}
& = & 16 \, \pi^2 \Im S
-{2 \over 3} \log \Im T \left \vert \vartheta_4 \left( T \right) \right\vert^4
-{2 \over 3} \log \Im U \left \vert \vartheta_4 \left( U \right) \right\vert^4
\nn \\
&& -b \log {\mu  \big/ M } \, ,
\label{htr}
\ea
where $S$ is the heterotic axion--dilaton field,
\be
\Im S  ={1 \over g^2_{\rm Het} } \, ,
\ee
and the fields $T$ and $U$ are respectively
the K\"{a}hler class and the complex structure moduli of the two-torus.
For the parametrization of Eq. (\ref{turr}),
the $\vartheta_4$ functions are obtained, as in Section 2,
when the $Z_2$ translation in $T^2$ acts as
a half-circumference translation, generated by the $(-1)^m$,
along $x_6$.
For the type I dual we have instead \cite{gk}:
\be
{16 \, \pi^2 \over g^2_{\rm   grav}(\mu )}
= 16 \, \pi^2 \Im S
- {2 \over 3} \log \Im U \left \vert \vartheta_4 \left( U \right) \right\vert^4
-b \log {\mu  \big/ M } \, .
\label{Itr}
\ee
In both the expressions (\ref{htr}) and (\ref{Itr})
we used the string scale $M \equiv
{1 / \sqrt{\alpha'}}$ and the infrared cut-off
$\mu $. These quantities can be either those of the heterotic 
or those of the type I string, their ratio being invariant under 
duality (cfr. Eq. (\ref{IIbthr})). 
On the type I side, $\Im S$ is associated to the coupling constant
of the gauge fields originating from the D9-branes sector and   
$U$ is still the complex structure modulus
of the untwisted two-torus. The heterotic volume modulus $T$,
on the other hand, is dual to $\tilde{S}^{\prime} \equiv -1 \big/ S^{\prime}$,
where $S^{\prime}$ is the type I field
whose VEV is the coupling constant of the gauge fields in the
D5-branes sector \cite{par}.
There is therefore an apparent mismatch between expression 
(\ref{htr}) and (\ref{Itr}). However, the functional dependence of
(\ref{htr}) on this modulus is such that,
for large $T$ (i.e. for small $S^{\prime}$), the dependence on this modulus 
disappears from the correction \footnote{More precisely, 
the string threshold diverges only logarithmically in $T$.}. 
On the other hand, the opposite limit, $T \to 0$, with $U$ fixed,
corresponds to a decompactification to a six 
dimensional ``true orbifold limit'', where the states with mass shifted by
the translation associated to the orbifold twist  become infinitely
massive. In this limit we expect therefore to recover
the properties of the orbifold discussed in \cite{gp}.  
In this limit, the correction (\ref{htr}) diverges linearly in
$\Im \tilde{T} \sim \Im S^{\prime}$, in agreement with the appearance
of a D5-branes sector. The limit in which the 
heterotic and type I models are comparable is therefore $T \to \infty$.
The full $T$-dependence shown in (\ref{htr}) has then to be
interpreted, on the type I side, as a non-perturbative correction.
In the limit $T \to 0$, namely the orbifold limit, 
the properties of the theory change dramatically, and the 
spectrum has to be found by expanding around a new perturbative
vacuum, namely that presented in Refs. \cite{gp}.

We can now proceed to compare all the constructions through the corrections
to the $R^2$ term. As we discussed, the two type I orientifolds,
$\Omega$ and $\tilde{\Omega}$, are trivially dual the one to the other.
The D9-branes sector of the $\Omega$ orientifold is therefore
mapped under duality into the D5-branes sector of the $\tilde{\Omega}$
construction. In particular, the coupling of the D9-branes
gauge fields is dual to the coupling of the D5-branes gauge fields.
This coupling is parametrized by the vacuum expectation value
of the imaginary part of a field, $S$, with:
\be
\Im S = {\rm e}^{- \phi_4} G^{1 / 4} \omega^2 \, .
\label{s}
\ee 
In the dual construction, the coupling of the D5-branes gauge fields
is parametrized by  the v.e.v. of the imaginary part of a field
$S^{\prime}$ (see also Refs. \cite{par}), with:
\be
\Im S^{\prime} = {\rm e}^{- \phi_4} G^{1 / 4} \omega^{-2} \, .
\label{sp}
\ee 
In (\ref{s}) and (\ref{sp}),
$\phi_4$ is the type I dilaton of four-dimensions, $\sqrt{G}$ the volume
of $T^2$ and $\omega^4$ the volume of the K3 ($\sim T^4 \big/ Z_2$)
of the type I compactification. 
Under duality, $S \leftrightarrow S^{\prime}$, as it can be seen also
by the comparison of (\ref{5branes}) and (\ref{Itr}).
From (\ref{IIbthr}) we can read then the asymptotic behavior
of the $\Omega$ orientifold in the small-$S$ limit. 
We learn that the contribution of this field vanishes in this limit,
signaling the breaking of S-duality.
By further comparison with the $R^2$ correction 
of the heterotic dual, given in (\ref{htr}), we can make an 
ansatz for the complete correction to the effective
coupling of this term.
By considering the ``vanishing'' of the contribution of any of
the fields $S$, $\tilde{T} \equiv -1 \big/ T$, 
$\tilde{U} \equiv -1 \big/ U$ in the small fields limits,
as opposed to the linear divergence when these fields are large,
we argue that the full non-perturbative extension of the gravitational 
correction (\ref{htr}) is given by the symmetrization of that expression in
the fields $\tilde{S}$, $T$, $U$ \footnote{To be more precise,
we should introduce here the fields $\tau_S \equiv 4 \pi S$
and $\tilde{\tau}_S \equiv \tau^{-1}_S$, and normalize 
the argument in the first term of the correction according to:
$\vartheta_4 \left( 6 \tilde{\tau}_S \right) $.  
However, for the sake of simplicity,
in this and in the subsequent formulae we don't care about
the normalization of the fields.}:
\ba
{16 \, \pi^2 \over g^2_{\rm   grav}}
 & \approx & 
-{2 \over 3} \log \Im \tilde{S} \vert \vartheta_4 ( \tilde{S} ) \vert^4
-{2 \over 3} \log \Im T \left \vert \vartheta_4 \left( T \right) \right\vert^4
-{2 \over 3} \log \Im U \left \vert \vartheta_4 \left( U \right) \right\vert^4
\nn \\
&& + {\cal E}\left( \tilde{S},T,U \right)
\, ,
\label{nptr}
\ea
where ${\cal E}\left( \tilde{S},T,U \right)$ stays for a series
of exponentials of the type ${\rm e}^{{\rm i}(k_1 \tilde{S}+k_2 T + k_3 U)}$,
in which the fields $\tilde{S}$, $T$, $U$ 
appear weighted by ``instanton numbers''. This term, 
symmetric in $\tilde{S}$, $T$, $U$, 
is suppressed in both the large and small fields limit\footnote{The 
limit in which these fields 
are small must in fact be taken after a  Poisson resummation, that converts 
this sum of exponentials into another sum of exponentials
of the inverse fields, still suppressed.}.
As a consequence, also the dependence on the field $S$, 
linearly divergent for large $\Im S$, is instead only logarithmic
when $S$ is small\footnote{We recall that under $-1 \big/ S \to S$, 
$\vartheta_4 ( -1 \big/ S) \to \vartheta_2 (S)$.}.
Note that if in Eq. (\ref{nptr}) we take the limit $\tilde{S} \to \infty$
(i.e. $S \to 0$), we obtain:
\be
{16 \, \pi^2 \over g^2_{\rm   grav}} \,
  \approx  \,
-{2 \over 3} \log \Im T \left \vert \vartheta_4 \left( T \right) \right\vert^4 
-{2 \over 3} \log \Im U \left \vert \vartheta_4 \left( U \right) \right\vert^4
+ {\cal O}\left( \log \Im S  \right)
\, .
\label{nptr2}
\ee
In this limit, the effective coupling indeed looks like the
non-perturbative expression of the coupling of an ${\cal N}_4=2$ theory 
with only D5-branes. We can in fact interpret, as usual,
the logarithmic dependence on the field, $S$, as
the remnant of non-perturbative effects, that
corrects the absence of any perturbative term.

The limit $\tilde{T} \equiv -1 \big/ T \to \infty$, $\tilde{T} < S$,
is a perturbative limit on the heterotic side. However,
as we argued, it should correspond to a limit of the theory in which
D5-branes massless states are expected to appear.
As long as $\Im \tilde{T} \ll \Im S$, these states are essentially
decoupled from the theory. As we approach the border of the heterotic 
perturbative regime, these states start to be dynamical,
and they cannot be neglected. Heterotic/type I duality in that
regime must be studied in the framework of the construction
of Refs. \cite{gp}: we argue that the appearance of a non-negligible 
$T$-dependence on the threshold corrections accounts for
the ``transition'' to this phase of the theory.
In the opposite limit, of ${\cal N}_4=4$ supersymmetry restoration,
this sector instead decouples.

For the ``winding breaking'' model, owing to the presence
of both D9- and D5-branes sectors,
the analogous of (\ref{Itr}) reads:
\be
{16 \, \pi^2 \over g^2_{\rm   grav}(\mu )}
= 16 \, \pi^2 \Im S + 16 \, \pi^2 \Im S^{\prime}
- {2 \over 3} \log \Im U \left \vert \vartheta_4 \left( U \right) \right\vert^4
-b \log {\mu  \big/ M } \, .
\ee
In this case, we cannot 
derive the strong coupling behavior of the threshold correction
by looking at an heterotic dual.
A naive investigation of the decompactification limit
in the $\tilde{\Omega}$ dual orientifold doesn't help either,
because, owing to the simultaneous presence of both
D9- and D5-branes, this is in any case a singular limit.
However, also in this model the ${\cal N}_4=2$
supersymmetry comes from the spontaneous breaking of ${\cal N}_4=4$,
which is restored in a corner of the moduli space.
In this phase, the D5-branes sector of the $\Omega$ orientifold
is expected to disappear, and the properties of the theory
should be analogous to those of the ${\cal N}_4=4$ phase of the
``momentum breaking'' model. 
We argue therefore that, as in that case, also here the
$S \to -1 \big/ S$ and  $S^{\prime} \to -1 \big/ S^{\prime}$ 
symmetries are broken, and the approximate behavior
of the $R^2$ threshold correction is as in (\ref{nptr}).

\noindent

\vskip 0.3cm
\setcounter{section}{4}
\setcounter{equation}{0}
\section*{\normalsize{\bf 4. Reduced rank type I models and duality}}

An analogous non-perturbative behavior is shown by other string
constructions, with a gauge group of reduced rank, $r=8$ and $r=4$.
Of these theories, we know only the type I, $\Omega$ orientifold
constructions, and their $\tilde{\Omega}$ duals.  
On the type I side, the reduction of the rank is obtained by
introducing a non-vanishing quantized antisymmetric NS-NS tensor $B_{ab}$.
Examples with ${\cal N}_4=4$ supersymmetry have been
constructed in Ref. \cite{bps,b}. This construction has then been extended
to ${\cal N}_4=2$, $Z_2$ non freely-acting orbifolds \cite{ka,a}.
Here we are interested in the case in which the $Z_2$
projection that partially breaks supersymmetry acts freely, 
as in Ref. \cite{adds}.
In this way, with a rank 2 antisymmetric tensor we obtain
a ``Scherk-Schwarz breaking'' model with maximal gauge group
$SO(16)$ and an ``M-theory breaking'' model with maximal gauge group
$SO(8)_{(9)} \times SO(8)_{(5)}$.
With an antisymmetric tensor of rank four we obtain then
a ``Scherk-Schwarz breaking'' model with maximal gauge group
$SO(8)$ and an ``M-theory breaking'' model with maximal gauge group
$SO(4)_{(9)} \times SO(4)_{(5)}$.
With appropriate Wilson lines, the above $SO(N)$ groups
can be reduced to the corresponding $U(N/2)$.
As for the rank 16 of the previous section, Wilson lines
are needed in order to introduce hypermultiplets in the 
``Scherk-Schwarz breaking'' models,
besides the four originating from the compact space.

The Klein Bottle, Annulus and M\"{o}bius strip amplitudes
for these models can be easily derived from the corresponding amplitudes
of the ``Scherk-Schwarz''-  and ``M-theory''-breaking 
${\cal N}_4=2$ models of Ref.\cite{adds}.
In our case, we work in four dimensions, and the shifted lattice sum
terms $Z_m$, defined in Ref. \cite{ads}, have to be adapted for a
two-dimensional lattice.
The models under consideration can be constructed also in five dimension,
because only one direction in the two-torus is shifted:
we can therefore always factorize the 
sum $\Gamma_{(1)}$ over the unshifted momenta 
of the lattice of one of the two circles. In the following, by $Z_m$ we will
actually mean $\Gamma_{(1)} \times Z_m$.
For the ``Scherk-Schwarz'' breaking, we have in this case
\footnote{For a definition of the characters $Q_0$, $Q_V$
we refer to \cite{adds}.}:
\be
{\cal K} = \sum_m
{1 \over 4} \left( Q_O + Q_V  \right) \left( q^2 \right) 
\left[ P Z_m + \tilde{W} (-1)^m Z_m \right] \, ,
\ee
where $q = {\rm e}^{- 2 \pi t}$, $P$ is the momentum sum of the 
lattice of $T^4$:
\be
P = \sum_m { q^{{\alpha' \over 2} m^T g^{-1} m}
\over \eta^4 \left( q^2  \right) } \, ,
\ee
and $\tilde{W}$ is the corresponding winding sum,
in which the windings satisfy the constraint imposed by 
a quantized non-vanishing antisymmetric tensor \cite{bps,b,a}:
\be
{2 \over \alpha'} B_{ab} n^b = 2 m_a \, ,
\ee
so that
\be
\tilde{W}=
2^{-4} \sum_{\epsilon=0,1} \sum_n
{ q^{ { 1 \over 2 \alpha'} n^T g n} {\rm e}^{ {2 i \pi \over \alpha'}
n^T B \epsilon} \over \eta^4 \left( q^2  \right) } \, , 
\ee
where the sum over $\epsilon$, a set of vectors with entries 0 or 1, 
introduces the projection along the directions of non-vanishing
$B_{ab}$.  
The Annulus amplitude reads:
\ba
{\cal A} & = & \sum_m \left\{ 
{ ( N_1 +N_2 )^2 \over 4} \left( Q_O +Q_V  \right) \left( \sqrt{q} \right) 
2^{r-4} \tilde{P} Z_m \right. \nn \\
&& \left. ~~~~~ -  
{ ( N_1 -N_2 )^2 \over 4} \left( Q_O -Q_V  \right) \left( \sqrt{q} \right)  
\left( { \vartheta^2_3 \vartheta^2_4 \over \eta^4 } \right)
\left( \sqrt{q} \right)
(-1)^m Z_m \right\} \, ,
\ea
where
\be
\tilde{P} \equiv
\sum_{\epsilon=0,1} \sum_m
{ q^{{\alpha' \over 2} (m + { 1 \over \alpha'} B \epsilon)^T g^{-1} 
(m+ { 1 \over \alpha'} B \epsilon)}
\over \eta^4 \left( \sqrt{q}  \right) } 
\ee
and $r$ is the rank of $B_{ab}$.
Finally, the M\"{o}bius strip amplitude is given by:
\ba
{\cal M} & = & - { ( N_1 +N_2 ) \over 4}
\sum_m \left\{ \left( \hat{Q}_O +\hat{Q}_V  \right) 
\left( -\sqrt{q} \right)
2^{(r-4) / 2} \hat{\tilde{P}} Z_m \, + \right. \nn \\
&& \left. ~~~~~ -  
\left( \hat{Q}_O -\hat{Q}_V  \right) \left( - \sqrt{q} \right) 
\left( { \hat{\vartheta}^2_3 \hat{\vartheta}^2_4 \over \hat{\eta}^4 } \right)
\left( -\sqrt{q} \right) (-1)^m Z_m \right\} \, ,
\ea
with
\be
\hat{\tilde{P}} \equiv
\sum_{\epsilon=0,1} \sum_m
{ q^{{\alpha' \over 2} (m + { 1 \over \alpha'} B \epsilon)^T g^{-1} 
(m+ { 1 \over \alpha'} B \epsilon)} \gamma_{\epsilon}
\over \hat{\eta}^4 \left( -\sqrt{q}  \right) } \, .
\ee
$\gamma_{\epsilon}$ is a cocycle, necessary in order to ensure
a correct particle interpretation (see \cite{a}).
Notice that the projection introduced by $B_{ab}$ acts on the
momenta of $T^4$, while the translation due to the 
supersymmetry breaking $Z^{\rm F}_2$ projection acts on the momenta of
the extra $T^2$.
The Chan--Paton factors are then fixed by the vanishing condition 
for the infrared divergences appearing in the so-called
``transverse channel'', in which the open string diagrams are
interpreted as diagrams for the  propagation of a closed string.
The proper ``time'' $\ell$ in the transverse channel is related to
the time $t$ in the direct channel in a different way for 
the three topologies, Klein bottle, Annulus and M\"{o}bius strip:
\ba
{\cal K} & : & \ell = {1 \over 2 t_{\cal K}} \, , \nn \\
{\cal A} & : & \ell = {2 \over t_{\cal A}} \,  ,\\
{\cal M} & : & \ell = {1 \over 2 t_{\cal M}} \, . \nn 
\ea
By transforming therefore the above amplitudes to the transverse channel
and looking at the divergence at the origin of the lattice 
($\ell \to \infty$), it is easy to see that the tadpole condition is modified:
with respect to the freely acting orbifold without antisymmetric tensor,
now the rank of the gauge group is reduced by a factor $r \big/ 2$.
In the transverse channel, after taking the $\ell \to \infty$ limit,
we obtain the following contributions
to the infrared divergence:
\ba
\tilde{{\cal K}}_0 & = & { 2^5 \over 4 \times 2} V_{(6)} \nn \\ 
&& \nn \\
\tilde{{\cal A}}_0 & = & { 2^{(r-4)-5} \over 4 \times 2} 
\left( N_1 + N_2 \right)^2 V_{(6)}       \nn \\
\tilde{{\cal M}}_0 & = & - { 2 \times 2^{(r-4)/2} \over 4 \times 2} 
\left( N_1 + N_2 \right) V_{(6)}  \nn
\ea
where
\be
V_{(6)} =
\sqrt{ \det \left( g_{(6)} \big/ \alpha^{\prime} \right)} .
\ee
They are canceled if:
\be
\left[  2^{r / 2} \left( N_1 + N_2 \right)- 2^5 \right]^2=0 \, ,
\ee
that means
\be
\left( N_1 + N_2 \right) = { 2^5 \over 2^{r / 2}} \, .
\ee
In the above amplitudes, the
splitting of the Chan--Paton factors into $N_1$ and $N_2$
is due to a Wilson line with entries ${1 \over 2}$, that can be chosen
such that $N_1=N_2=N=32 \big/ 2^{r / 2}$.
In this case, the vector multiplets are in the
$\left( {\bf { N ( N-1) \big/ 2}}, {\bf {1}} \right)$ $\oplus$
$\left( {\bf {1}}, {\bf { N ( N-1) \big/ 2}} \right)$ adjoint representations
of $SO(N) \times SO(N)$, the hypermultiplets in the bifundamental
$( {\bf {N}}, {\bf {N}} )$.
The case of our interest is obtained
when a further Wilson line is introduced, with entries ${1 \over 4}$,
that breaks the gauge group to its unitary subgroup, 
$U(N / 2) \times U(N / 2)$.

In the case of ``M-theory'' breaking, the amplitudes are, in the
simplest case without Wilson lines:
\be
{\cal K} = \sum_m \left\{ {1 \over 4} 
\left( Q_O + Q_V \right) \left( q^2 \right) ( P + \tilde{W} ) Z_m 
+ { 2^{( 4 - r) / 2}   \over 2} \left( Q_S + Q_C \right) \left( q^2 \right)
\left( { \vartheta^2_2 \vartheta^2_3 \over \eta^4 } \right) 
\left( q^2 \right) Z_{m+{1 \over 2}} \right\} \, ,
\ee
\ba
{\cal A} & = & {1 \over 4} \sum_m \left\{  \left[  
\left( Q_O + Q_V  \right) \left( \sqrt{q}  \right)
\left( 2^{r - 4} n^2 \tilde{P} + d^2 W \right) \right] Z_{2m} \right. \nn \\
&& \left. ~~~~~ + 
{ 2^{r / 2} \over 2} (n d) \left( Q_S + Q_C \right)
\left( \sqrt{q}  \right)
\left( { \vartheta^2_2 \vartheta^2_3 \over  \eta^4 } \right) 
\left( \sqrt{q}  \right) \right\} Z_{2m+1} \, ,
\ea
\ba
{\cal M} & = & - { 1 \over 4} \sum_m \left\{
\left( \hat{Q}_O + \hat{Q}_V  \right) \left( - \sqrt{q}  \right)
\left[ 2^{(r-4)/2} n \hat{\tilde{P}} +2^{-2} d  \hat{\tilde{W}} \right] 
Z_{4m} \right.
\nn \\
&& ~~~~~ \left. -
(n+d) \left( \hat{Q}_O - \hat{Q}_V  \right) \left( - \sqrt{q}  \right)  
\left( { \hat{\vartheta}^2_3 \hat{\vartheta}^2_4 \over  \hat{\eta}^4 } \right) 
\left( -\sqrt{q}  \right) Z_{4m+2} \right\} \, ,
\ea
where
\be
\hat{\tilde{W}} = \sum_{\epsilon=0,1} \sum_n
{ q^{ {1 \over 2 \alpha^{\prime}} n^{\rm T} g n}
{\rm e}^{ {2 i \pi \over \alpha^{\prime}} n^{\rm T} B \epsilon }
\tilde{\gamma}_{\epsilon} \over
\hat{\eta}^4 \left( - \sqrt{q} \right) } \, ,
\ee
with $\tilde{\gamma}_{\epsilon}$ a further, independent cocycle
(see Ref. \cite{a}).
In this case, the further reduction of the rank of the gauge group
involves the parts coming both from the D9-branes and the D5-branes
sectors, in a symmetric way. 
The contributions to the infrared divergence are: 
\ba
\tilde{\cal K}_0 & = & {2^5 \over 4 \times 2} V_{(2)} 
\left[ V_{(4)} + { 2^{-r} \over V_{(4)}} \right] \nn \\
\tilde{\cal A}_0 & = & {2^{-5} \over 4 \times 2}  V_{(2)} 
\left[ V_{(4)} \, 2^r \, n^2 \; 
+ \, { d^2 \over V_{(4)}} \right] \nn \\
\tilde{\cal M}_0 & = & - {2 \over 4 \times 2}  V_{(2)} 
\left[ V_{(4)} \, 2^{r / 2} \, n \; 
+ \, { 2^{- r / 2} \, d^2 \over V_{(4)}} \right] \, ,
\ea
where $V_{(2)}=\sqrt{\det \left(g_{(2)} \big/ \alpha^{\prime} \right)}$ and
$V_{(4)}=\sqrt{\det \left(g_{(4)} \big/ \alpha^{\prime} \right)}$ are
the volumes of  the two- and the four-torus 
respectively. The tadpole cancellation conditions are:
\be
n={2^5 \over 2^{r / 2}}, ~~~~~~~
d={2^5 \over 2^{r / 2}}.
\ee

The construction of the $\tilde{\Omega}$ orientifolds
is analogous, and their partition functions
are obtained by those of the ordinary orientifolds, by
exchanging the role of D9- and D5-branes.
The description from the bulk viewpoint, suitable
to investigate the decompactification limit, is then obtained 
by shifting all the windings and doubling all the momenta of the coordinates 
transverse to the D5-branes in the Klein bottle and in the open string 
diagrams. As before, Wilson lines are needed in order to break the
gauge group to the Abelian subgroup, and obtain duality
between type I gauge groups and the $\tilde{\Omega}$ constructions
with the D5-branes sitting on the corresponding O5 orientifold plane,
in order to have a local cancellation of the tadpoles.
In this case, however,
we don't have the heterotic dual orbifold constructions:
the only heterotic orbifolds with reduced rank of the gauge group
in which the gravitational corrections
have a perturbative expression like (\ref{htr}), with a factorization
of the moduli $T$ and $U$, are those considered in Refs. \cite{gkp}
\footnote{Heterotic constructions with gauge groups of reduced rank
have also been considered in Ref. \cite{chl},
and their relation to type I constructions in Refs. \cite{bps,b,w}.}:
they are necessarily constructed at a point in the moduli space
in which the higher level (= reduced rank) gauge group 
is realized with real fermions, the points 
$\left[ SU(2)_{\kappa} \right]^{16/ \kappa}= 
\left[ SO(3)_{\kappa / 2} \right]^{16/ \kappa}$. 
In these heterotic orbifolds, 
the massless states originating from the currents 
have an effective ${\cal N}_4=4$ supersymmetry.  
They are dual to type IIA orbifolds with spontaneously broken
${\cal N}_4=4$ supersymmetry, that are singular limits in the moduli space of 
K3 fibrations. The full, non-perturbative gravitational correction 
can be computed perturbatively on the type IIA side (see Refs. \cite{gkp}), 
and it turns out that there is no limit in the space of moduli
$S$, $T$, $U$ in which there is a restoration of the ${\cal N}_4=8$ 
supersymmetry.

Even without the heterotic duals, from which we would learn
something more about one of the moduli,
the comparison of ``$\Omega$'' and ``$\tilde{\Omega}$'' constructions
allows us to see the spontaneous breaking of the ${\cal N}_4=8$
supersymmetry, as it was for the previous case with gauge group
of rank 16. By looking at the various limits of these theories,
with and without either D9- or D5-branes,
we conclude that the gravitational corrections have an expression
analogous to (\ref{nptr}).

\subsection*{\sl A note on the effective gauge coupling}

The rank reduction indeed corresponds to a raising of the
level of the algebra of the gauge group.
In order to define the level, we don't need in fact to work
with an explicit realization via Ka\v{c}-Moody characters:
the level can be intrinsically defined also through the strength
of the coupling of the gauge group.
Since the rank is halved when two isomorphic factors of the gauge group
are identified (this is the effect of the operation that
leaves massless the states invariant under the exchange of these factors 
and lifts the mass of the states which are odd),
the level is doubled. The relation between level $\kappa$
and level 1 coupling is:
\be
{1 \over g^2_{\kappa}} = {\kappa \over g^2_{( \kappa = 1)}} \propto 
{1 \over g^2_{(\kappa =1)} N} \, ,
\label{gn}
\ee
where $N$ is the ``effective'' number of D9- or D5-branes, namely
the number of branes modulo $Z_2$, level-raising identifications.
Raising the level corresponds therefore to weakening the effective 
gauge coupling $g_{YM} \equiv g_{\kappa} \propto 
g_{(\kappa =1)} \sqrt{N}$. 
If we now extrapolate from Eq. (\ref{gn}) and treat $\kappa$ and $N$
as parameters that can take any integer value, we can consider
arbitrary high levels. 
This leads to a decoupling of the Yang Mills fields, that
tend to be frozen to constant values: this is the no branes
limit, with no gauge group.
In the opposite limit, $N \to \infty$,
$g_{YM}$ is very big, and the Yang Mills fields  
dominate over the other fields: this is the limit of decoupling of gravity,
whose coupling is $g \equiv g_{(\kappa =1)}$.
We can look at this limit in another, equivalent way:
we can rescale the couplings is such a way that
the effective gauge coupling remains fixed. In the large $N$ limit
we have therefore $g \to 0$, with
$g^2 N$ fixed. This is precisely the 't~Hooft limit;
in the literature, what we call the effective level--$\kappa$
gauge coupling is indicated by $\Lambda_{QCD}$.
In the theories we are considering, there are 
D9- and D5-branes, with different four-dimensional coupling.
The above decoupling argument applies separately for both the D9-
and D5-branes sectors.
When both the types of branes are present, the effective coupling of
the bulk fields is in fact a combination of the two: 
$g^{-2}_{\rm bulk} \sim g^{-2}_{(9)} + g^{-2}_{(5)} $,
and the effective couplings of the two parts of the gauge group
are $g_{YM (9)} \propto g_{(9)} \sqrt{N}$, $g_{YM (5)} \propto
g_{(5)} \sqrt{N}$.

\noindent

\vskip 0.3cm
\setcounter{section}{5}
\setcounter{equation}{0}
\section*{\normalsize{\bf 5. Conclusions}}

In this work we have investigated the non-perturbative behavior
of a class of four dimensional string models with ${\cal N}_4=2$
supersymmetry, in which there is a spontaneous breaking of 
${\cal N}_4=8$. Some of these theories have dual realizations
as freely acting orbifolds of both the heterotic and the type I string,
others have only a type I orbifold realization.
The spontaneous breaking of the ${\cal N}_4=8$ supersymmetry is
due to a super-Higgs mechanism, non-perturbative in all
the dual constructions: the supersymmetry restoration can be
observed by comparing the heterotic and type I effective
theories with the duals constructed as freely acting orbifolds of the
type IIB string orientifolded with $\tilde{\Omega}$,
the product of the world-sheet parity times the target-space
reflection along four coordinates, compactified on $T^4$.  
In these constructions, the role of D9- and D5-branes is exchanged,
as compared to the ordinary type I orientifolds.
Owing to the free action of the orbifold projection, these theories
possess phases in which only D5-branes appear.
In these configurations, it is possible to investigate
the limit in which the space transverse to the D5-branes
is decompactified.
Owing to the breaking of T-duality associated to the
supersymmetry breaking projection produced by the D5-branes,
this is a limit in which, from the bulk point of view, namely
at large distance from the D-branes, there is an
approximate restoration of the maximal amount of supersymmetry.

We have compared the dual constructions by looking at the
renormalization of the effective coupling of the $R^2$ term,
that depends on the moduli of the vector manifold, and therefore
on the heterotic dilaton--axion field and on the type I
vector coupling fields, ``$S$'' and ``$S^{\prime}$''.
Since the ${\cal N}_4=4$ heterotic string possesses S-duality \cite{sen},
the existence of a spontaneous breaking of ${\cal N}_4=8$,
that can be restored at the strong coupling,
means that, in this case, the projection that breaks the
${\cal N}_4=4$ supersymmetry to ${\cal N}_4=2$ has an
action also on $S$, the heterotic dilaton--axion field.
 
For simplicity, we have limited the analysis of these theories
to the region of the moduli space in which the gauge group is 
broken to its Abelian subgroup. However, the ``decompactification'' arguments
that led us to conclusions about their non-perturbative behavior
can be applied also in regions in which the gauge group is non-Abelian.
The key point is that the gauge group
must correspond to a choice of Wilson lines dualizable to a Dp-branes
configuration, p $<$ 9, in which there is a local cancellation
of the orientifold charge, without dilaton gradient along some transverse
directions.

\vskip 1.cm
\centerline{\bf Acknowledgements}
\noindent
I thank C. Angelantonj, C. Bachas, K. Benakli, C. Kounnas, Y. Oz, 
P. Mayr and A. Zaffaroni for valuable discussions,
the \'Ecole Normale of Paris for kind hospitality,
the \'Ecole Polytechnique and the CNRS for kind hospitality and financial 
support.
I want to thank also J.P. Derendinger for interesting discussions and
hospitality in the University of Neuch\^{a}tel, where part of this work was 
carried out.
This work was partially supported also by the Swiss National Science
Foundation {and the Swiss Office for Education and Science.

\newpage 

\vskip 0.3cm
\setcounter{section}{0}
\setcounter{equation}{0}
\renewcommand{\theequation}{A. \arabic{equation}}
\section*{\normalsize{\centerline{\bf Appendix:
The transverse channels for the reduced rank type I models}}}
\noindent
We quote here the Klein Bottle, Annulus
and M\"{o}bius Strip contributions in the transverse channel
for the reduced rank type I models of Section 4.
For the rank reductions of the Scherk-Schwarz model of Ref. \cite{adds},
we have: 
\ba
\tilde{\cal K} & = & {2^5 \over 4} \left( Q_O + Q_V  \right) 
\left( {\rm i} \ell  \right)
\left[ \, V_{(4)} \,
\sum_n { \left( {\rm e}^{- 2 \pi \ell} \right)^{ { 1 \over \alpha^{\prime}}
n^{\rm T} g n} \over \eta^4 \left({\rm i} \ell  \right) } 
\, V_{(2)} \, \left( \sum_{n^{\prime}}  \tilde{Z}_{2n^{\prime}} \right) 
\, + \right. \nn \\
&& ~~\left. + \,
{ 2^{-4} \over V_{(4)} } \, \sum_{\epsilon=0,1} \sum_m
{ \left( {\rm e}^{ -2 \pi \ell} \right)^{\alpha^{\prime} ( m + {1 \over 
\alpha^{\prime}} B \epsilon)^{\rm T} g^{-1}  
( m + {1 \over \alpha^{\prime}} B \epsilon)} \over
\eta^4 \left( {\rm i} \ell  \right) }   
V_{(2)} \, \left( \sum_{n^{\prime}}  \tilde{Z}_{2n^{\prime}+1} \right) 
\right] \, ,
\ea
\ba
\tilde{\cal A} & = & { 2^{-5} \over 4} \left\{ 
\left( Q_O + Q_V  \right) \left( {\rm i} \ell  \right) 
2^{r - 4} \, V_{(4)} \, \left( N^2_1 + N^2_2 \right)\, 
\sum_{\epsilon=0,1}
\sum_n { \left( {\rm e}^{- 2 \pi \ell} \right)^{ { 1 \over 4 \alpha^{\prime}}
n^{\rm T} g n} {\rm e}^{ { 2 {\rm i} \pi \over \alpha^{\prime} }
n^{\rm T} B \epsilon}  
\over \eta^4 \left( {\rm i} \ell  \right) } \, \times \right. \nn \\
&& ~~~~~~~~~~~~~~~~~~~~~~~~~~~~\left. \times  \, 
V_{(2)} \, \left( \sum_{n^{\prime}} \tilde{Z}_{2n^{\prime}}  \right)
\right.  \, + \nn \\
&& \left.  
+ \, 4 \, (N^2_1 - N^2_2 ) \,
\left( Q_S + Q_C  \right) \left( {\rm i} \ell  \right) 
\left( { \vartheta^2_2 \vartheta^2_3 \over \eta^4 } \right)
\left( {\rm i} \ell  \right) \,
V_{(2)} \, \left( \sum_{n^{\prime}} 
\tilde{Z}_{2n^{\prime}+1}  \right)
\right\} \, ,
\ea
\ba
\tilde{\cal M} & = & - {2 \over 4} \left( N_1 + N_2  \right) \left\{   
\left( \hat{Q}_O + \hat{Q}_V \right)
\left( {\rm i} \ell + {1 \over 2} \right) \,
2^{ (r - 4) / 2} \, V_{(4)} \, \times \right. \nn \\
&& ~~~~~~~~~~~~~~~ \left. \times \, 
\sum_{\epsilon=0,1}
\sum_n { \left( {\rm e}^{- 2 \pi \ell} \right)^{ { 1 \over  \alpha^{\prime}}
n^{\rm T} g n} {\rm e}^{ { 2 {\rm i} \pi \over \alpha^{\prime} }
n^{\rm T} B \epsilon}  \gamma_{\epsilon}
\over \hat{\eta}^4 \left( {\rm i} \ell + {1 \over 2} \right) } \,
V_{(2)} \, \left( \sum_{n^{\prime}} 
\tilde{Z}_{2n^{\prime}}  \right) \, + 
\right.  \nn \\
&& ~~~ \left.  + \,
\left( \hat{Q}_O - \hat{Q}_V \right)
\left( {\rm i} \ell + {1 \over 2} \right)
\left( { \hat{\vartheta}^2_3 \hat{\vartheta}^2_4 \over \hat{\eta}^4 } \right)
\left( {\rm i} \ell  + {1 \over 2} \right) 
V_{(2)} \, \left( \sum_{n^{\prime}} 
\tilde{Z}_{2n^{\prime}+1}  \right) \right\} \, . \nn \\
&&
\ea

For the ``M-theory breaking'' model we have:
\ba
\tilde{\cal K} & = & {2^5 \over 4} \left( Q_O + Q_V  \right) 
\left( {\rm i} \ell  \right)
\left[ \, V_{(4)} \,
\sum_n { \left( {\rm e}^{- 2 \pi \ell} \right)^{ { 1 \over \alpha^{\prime}}
n^{\rm T} g n} \over \eta^4 \left({\rm i} \ell  \right) } \, + \right. \nn \\
&& ~~\left. + \,
{ 2^{-4} \over V_{(4)} } \, \sum_{\epsilon=0,1} \sum_m
{ \left( {\rm e}^{ -2 \pi \ell} \right)^{\alpha^{\prime} ( m + {1 \over 
\alpha^{\prime}} B \epsilon)^{\rm T} g^{-1}  
( m + {1 \over \alpha^{\prime}} B \epsilon)} \over
\eta^4 \left( {\rm i} \ell  \right) }  \right] 
\, V_{(2)} \, \left( \sum_{n^{\prime}}  \tilde{Z}_{2n^{\prime}} \right) \, 
+ \nn \\
&& + \,
{ 2^{5 - r / 2} \over 2} \left( Q_O - Q_V  \right) \left( {\rm i} \ell  \right)
\left( \vartheta^2_3 \vartheta^2_4 \over \eta^4  \right) 
\left( {\rm i} \ell  \right) \,
V_{(2)} \, \left( \sum_{n^{\prime}}  (-1)^{n^{\prime}} 
\tilde{Z}_{2n^{\prime}} \right) \, ,  
\ea
\ba
\tilde{\cal A} & = & { 2^{-5} \over 4} \left\{ 
\left( Q_O + Q_V  \right) \left( {\rm i} \ell  \right) 
\left[ 2^{r - 4} \, V_{(4)} \, n^2 \, \sum_{\epsilon=0,1}
\sum_n { \left( {\rm e}^{- 2 \pi \ell} \right)^{ { 1 \over 4 \alpha^{\prime}}
n^{\rm T} g n} {\rm e}^{ { 2 {\rm i} \pi \over \alpha^{\prime} }
n^{\rm T} B \epsilon}  
\over \eta^4 \left( {\rm i} \ell  \right) }  \right. \right. \, + \nn \\
&& \left.  \left. + \,{ 1 \over V_{(4)}} \, d^2 \, 
{ \left( {\rm e}^{ -2 \pi \ell} \right)^{{ \alpha^{\prime} \over 4}
m^{\rm T} g^{-1} m } \over
\eta^4 \left( {\rm i} \ell  \right) } \right]
V_{(2)} \, \left( \sum_{n^{\prime}} \tilde{Z}_{2n^{\prime}}  \right)
\right. \, + \nn \\
&& \left.
+  2 \times 2^{r / 2} \, (n d ) \,
\left( Q_O - Q_V  \right) \left( {\rm i} \ell  \right) 
\left( { \vartheta^2_3 \vartheta^2_4 \over \eta^4 } \right)
\left( {\rm i} \ell  \right) \,
V_{(2)} \, \left( \sum_{n^{\prime}} (-1)^{n^{\prime}}
\tilde{Z}_{2n^{\prime}}  \right) \right\} \, , \nn \\
&&
\ea
\ba
\tilde{\cal M} & = & - {2 \over 4} \left\{   
\left( \hat{Q}_O + \hat{Q}_V \right)
\left( {\rm i} \ell + {1 \over 2} \right)
\left[ 2^{ (r - 4) / 2} \, V_{(4)} \, n \,  
\sum_{\epsilon=0,1}
\sum_n { \left( {\rm e}^{- 2 \pi \ell} \right)^{ { 1 \over  \alpha^{\prime}}
n^{\rm T} g n} {\rm e}^{ { 2 {\rm i} \pi \over \alpha^{\prime} }
n^{\rm T} B \epsilon}  \gamma_{\epsilon}
\over \hat{\eta}^4 \left( {\rm i} \ell + {1 \over 2} \right) } \, + 
\right. \right. \nn \\
&& ~~~\left.  \left.
+ { 2^{-2} \over V_{(4)}} \, d \, 
\sum_{\epsilon=0,1} \sum_m
{ \left( {\rm e}^{ -2 \pi \ell} \right)^{\alpha^{\prime} ( m + {1 \over 
\alpha^{\prime}} B \epsilon)^{\rm T} g^{-1}  
( m + {1 \over \alpha^{\prime}} B \epsilon)} \tilde{\gamma}_{\epsilon} \over
\hat{\eta}^4 \left( {\rm i} \ell + {1 \over 2} \right) }
\right] \, V_{(2)} \, \left( \sum_{n^{\prime}} 
\tilde{Z}_{2n^{\prime}}  \right) \, + \right. \nn \\
&& \left. \, + \,
\left( \hat{Q}_O - \hat{Q}_V \right)
\left( {\rm i} \ell + {1 \over 2} \right)
\left( { \hat{\vartheta}^2_3 \hat{\vartheta}^2_4 \over \hat{\eta}^4 } \right)
\left( {\rm i} \ell  + {1 \over 2} \right) \, (n+d) \,
V_{(2)} \, \left( \sum_{n^{\prime}} (-1)^{n^{\prime}}
\tilde{Z}_{2n^{\prime}}  \right) \right\} \, , \nn \\
&&
\ea
where
\be
V_{(4)} = \sqrt{\det \left(g_{(4)} \big/ \alpha^{\prime} \right)}
, ~~~~~~~~
V_{(2)} = \sqrt{\det \left(g_{(2)} \big/ \alpha^{\prime} \right)}
\ee
are the volume of $T^4$ and $T^2$ respectively.

\end{document}